\documentclass[11pt,a4paper]{article}
\usepackage{amsmath,amssymb}
\usepackage{amsfonts}
\usepackage{amsthm}
\usepackage{cite}
\theoremstyle{plain}
\newtheorem{theorem}{Theorem}

\theoremstyle{definition}
\newtheorem{definition}{Definition}
\newtheorem{assumption}{Assumption}

\newcommand{\Tr}{\operatorname{Tr}}
\newcommand{\rmd}{\mathrm{d}}

\newcommand{\rmi}{\mathrm{i}}
\newcommand{\ent}{\boldsymbol{|}}
\newcommand{\norm}[1]{\left\Vert#1\right\Vert}

\newcommand{\openone}{{\mathchoice \mathrm{1\mskip-4mu l} \mathrm{1\mskip-4mu l}
\mathrm{1\mskip-4.5mu l} \mathrm{1\mskip-5mu l}}}
\newcommand{\E}{\operatorname{\mathbb{E}}}

\newcommand{\rmq}{\mathrm{q}}
\newcommand{\rmc}{\mathrm{c}}

\newcommand{\calY}{\mathcal{Y}}
\newcommand{\calH}{\mathcal{H}}
\newcommand{\calL}{\mathcal{L}}
\newcommand{\calT}{\mathcal{T}}

\newcommand{\calF}{\mathcal{F}}
\newcommand{\calA}{\mathcal{A}}
\newcommand{\calB}{\mathcal{B}}
\newcommand{\calM}{\mathcal{M}}
\newcommand{\calS}{\mathcal{S}}
\newcommand{\calU}{\mathcal{U}}
\newcommand{\calX}{\mathcal{X}}
\begin{document}

\title{Entropic bounds and continual measurements}

\author{Alberto Barchielli
\\
{ \small Politecnico di Milano, Dipartimento di Matematica,}
\\
{\small Piazza Leonardo da Vinci 32, I-20133 Milano, Italy.}
\\
{\small E-mail: Alberto.Barchielli@polimi.it}
\\ {}
\\
Giancarlo Lupieri
\\
{ \small Universit\`a degli Studi di Milano, Dipartimento di Fisica,}
\\  {\small Via Celoria
16, I-20133 Milano, Italy.}
\\ {\small E-mail: Giancarlo.Lupieri@mi.infn.it}}

\date{November 3, 2005}

\maketitle

\abstract{Some bounds on the entropic informational quantities related to a quantum continual
measurement are obtained and the time dependencies of these quantities are studied.}

\section{Introduction}

In the problem of information transmission through quantum systems, various entropic
quantities appear which characterize the performances of the encoding and decoding
apparatuses. Due to the peculiar character of a quantum measurement, many bounds on the
informational quantities involved have been proved to hold
\cite{Gro71,Lin72,Hol73,Oza86,SchWW96,BarL05os,BarL05QP,BarL05qic}. In the case of
measurements continual in time, these bounds acquire new aspects (family of measurements are
now involved) and new problems arise. A typical question is about which of the various
entropic measures of information is monotonically increasing or decreasing in time. We already
started the study of this subject in Refs.\ \cite{Bar01,BarL04}; here we apply to the case of
continual measurements the new techniques developed \cite{BarL05os,BarL05QP,BarL05qic} for the
time independent case.

\subsection{Notations and preliminaries}

We denote by $\calL(\calA;\calB)$ the space of bounded linear operators from $\calA$ to
$\calB$, where $\calA,\,\calB$ are Banach spaces; moreover we set
$\calL(\calA):=\calL(\calA;\calA)$.

Let $\calH$ be a separable complex Hilbert space; a normal state on $\calL(\calH)$ is
identified with a statistical operator, $\calT(\calH)$ and $\calS(\calH)\subset \calT(\calH)$
are the trace-class and the space of the statistical operators on $\calH$, respectively, and
$\norm{\rho}_1:= \Tr \sqrt{\rho^*\rho}$, $\langle \rho,a\rangle := \Tr_{\calH} \{\rho a\}$,
$\rho \in \calT(\calH)$, $a\in \calL(\calH)$.

More generally, if $a$ belongs to a $W^*$-algebra and $\rho$ to its dual $\calM^*$ or predual
$\calM_*$, the functional $\rho$ applied to $a$ is denoted by $\langle \rho,a\rangle$.

\subsubsection{A quantum/classical algebra} \label{qcalgebra}
Let $(\Omega,\calF,Q)$ be a measure space, where $Q$ is a $\sigma$-finite measure. By Theorem
1.22.13 of \cite{Sak71}, the $W^*$-algebra $L^\infty(\Omega,\calF,Q)\otimes \calL(\calH)$
($W^*$-tensor product) is naturally isomorphic to the $W^*$-algebra
$L^\infty\big(\Omega,\calF,Q;\calL(\calH)\big)$ of all the $\calL(\calH)$-valued
$Q$-essentially bounded weakly$^*$ measurable functions on $\Omega$. Moreover (\cite{Sak71},
Proposition 1.22.12), the predual of this $W^*$-algebra is
$L^1\big(\Omega,\calF,Q;\calT(\calH)\big)$, the Banach space of all the $\calT(\calH)$-valued
Bochner $Q$-integrable functions on $\Omega$, and this predual is naturally isomorphic to
$L^1(\Omega,\calF,Q)\otimes \calT(\calH)$ (tensor product with respect to the greatest cross
norm --- \cite{Sak71}, pp.\ 45, 58, 59, 67, 68).

Let us note that a normal state $\sigma$ on $L^\infty\big(\Omega,\calF,Q;\calL(\calH)\big)$ is
a measurable function $\omega \mapsto \sigma(\omega)\in \calT(\calH)$, $\sigma(\omega)\geq 0$,
such that $\Tr_{\calH}\{\sigma(\omega)\}$ is a probability density with respect to $Q$.

\subsection{Quantum channels and entropies}

\subsubsection{Relative and mutual entropies}

The general definition of the relative entropy $S(\Sigma|\Pi)$ for two states $\Sigma$ and
$\Pi$ is given in \cite{OhyP93}; here we give only some particular cases of the general
definition.

Let us consider two quantum states $ \sigma,\,\tau\in \calS(\calH)$ and two classical states
$q_k$ on $L^\infty(\Omega,\calF,Q)$ (two probability densities with respect to $Q$). The
quantum relative entropy and the classical one are
\begin{subequations}
\begin{gather}\label{relqentropy}
S_\rmq(\sigma|\tau)=\Tr_\calH\{\sigma(\log \sigma-\log \tau)\},
\\
S_\rmc(q_1|q_2)= \int_\Omega Q(\rmd \omega)\,q_1(\omega) \log\frac{q_1(\omega)}{q_2( \omega)}
\,.
\end{gather}
\end{subequations}

We shall need also the von Neumann entropy of a state $\tau\in \calS(\calH)$: $S_\rmq(\tau):=
- \Tr \{\tau \log \tau\}$.

Let us consider now two normal states $\sigma_k$  on
$L^\infty\big(\Omega,\calF,Q;\calL(\calH)\big)$ and set $q_k(\omega):= \Tr
\{\sigma_k(\omega)\}$, $\varrho_k(\omega):= \sigma_k(\omega)/q_k(\omega)$ (these definitions
hold where the denominators do not vanish and are completed arbitrarily where the denominators
vanish). Then, the relative entropy is
\begin{subequations}
\begin{align}\label{cqS1}
S(\sigma_1|\sigma_2) &= \int_\Omega Q(\rmd \omega) \Tr_\calH\left\{\sigma_1(\omega)\big(\log
\sigma_1(\omega)-\log \sigma_2(\omega)\big)\right\}
\\ \label{cqS2}
{}&= S_\rmc(q_1|q_2) + \int_\Omega Q(\rmd \omega) \, q_1(\omega)S_\rmq\big(\varrho_1(\omega)|
\varrho_2(\omega)\big).
\end{align}
\end{subequations}
We are using a subscript ``c'' for classical entropies, a subscript ``q'' for purely quantum
ones and no subscript for general entropies, eventually of a mixed character.

Classically a mutual entropy is the relative entropy of a joint probability with respect to
the product of its marginals and this key notion can be generalized immediately to states on
von Neumann algebras, every times we have a state on a tensor product of algebras
\cite{BarL05os,BarL05QP,BarL05qic}.

\subsubsection{Channels}
\begin{definition}
(\cite{OhyP93} p.\ 137) Let $\calM_1$ and $\calM_2$ be two $W^*$-algebras. A linear map
$\Lambda^*$ from $\calM_2$ to $\calM_1$ is said to be a \emph{channel} if it is completely
positive, unital (i.e.\ identity preserving) and normal (or, equivalently, weakly$^*$
continuous).
\end{definition}

Due to the equivalence \cite{Dix57} of w$^*$-continuity and existence of a preadjoint
$\Lambda$, a \emph{channel} is equivalently defined by: $\Lambda$ is a completely positive
linear map from the predual $\calM_{1*}$ to the predual $\calM_{2*}$, normalized in the sense
that $\langle \Lambda[\rho], \openone_2 \rangle_2= \langle \rho,\openone_1\rangle_1$, $\forall
\rho\in \calM_{1*}$. Let us note also that $\Lambda$ maps normal states on $\calM_1$ into
normal states on $\calM_2$.

A key result which follows from the convexity properties of the relative entropy is
\emph{Uhlmann monotonicity theorem} (\cite{OhyP93}, Theor.\ 1.5 p.\ 21), which implies that
channels decrease the relative entropy.
\begin{theorem}\label{Uhltheo}
If $\Sigma$ and $\Pi$ are two normal states on $\calM_1$ and $\Lambda^*$ is a channel from
$\calM_2\to \calM_1$, then $ S(\Sigma| \Pi)\geq S(\Lambda[\Sigma]| \Lambda[\Pi])$.
\end{theorem}

\subsection{Continual measurements}
Let us axiomatize the properties of a probability space where an independent-increment process
lives and that ones of the $\sigma$-algebras generated by its increments. The probability
measure $Q_1$ we are introducing will play the role of a reference measure.

\begin{assumption}\label{Ass1}
Let $(X,\calX,Q_1)$ be a probability space with $(X,\calX)$ standard Borel. Moreover:
\begin{enumerate}
\item $\{\calX^s_t, 0\leq s\leq t\}$ is a two-times filtration of sub-$\sigma$-algebras:
$\calX^s_t\subset \calX^r_T \subset \calX$ for $0\leq r \leq s\leq t\leq T$;
\item
$\forall t\geq 0$, $\calX^t_t$ is trivial;
\item
$\displaystyle \calX^s_t= \bigwedge_{T: T>t} \calX^s_T$ for $0\leq s\leq t$;
\item $\displaystyle \calX^s_t= \bigvee_{r: s<r<t} \calX^r_t$ for $0\leq s< t$;
\item $\displaystyle \calX= \bigvee_{t: T>0} \calX^0_t$;
\item
for $0\leq r \leq s \leq t \leq T$, $\calX^r_s$ and $\calX^t_T$ are $Q_1$-independent.
\end{enumerate}
\end{assumption}

Continual measurements are a quantum analog of classical processes with independent increments
\cite{Hol01,BarL04}. As any kind of quantum measurement, a continual measurement is
represented by \emph{instruments} \cite{Dav76,Oza84,Oza85a}, but, as shown in \cite{BarL05QP},
instruments are equivalent to particular types of channels. Here we introduce continual
measurements directly as a family of channels satisfying a set of axioms (cf.\ also
\cite{Bar86,BarL04}).

\begin{assumption}\label{Ass2}
Let $\calH$ be a separable complex Hilbert space. For all $s,t$, $0\leq s \leq t$, we have a
channel
\[
\tilde\Lambda^s_t : L^1 \big(X,\calX^0_s, Q_1; \calT(\calH)\big) \to L^1 \big(X,\calX^0_t,
Q_1; \calT(\calH)\big)
\]
such that
\begin{enumerate}
\item
$\tilde\Lambda^t_t=\openone$, \ $t \geq 0$;
\item
$\tilde\Lambda^s_t\circ \tilde\Lambda^r_s= \tilde\Lambda^r_t$, \ $0\leq r\leq s\leq t$;
\item
$\forall \eta\in \calT(\calH)$, \ $\tilde\Lambda^s_t[\eta]$ is $\calX^s_t$-measurable, \
$0\leq s\leq t$;
\item
$\forall \eta\in \calT(\calH)$, $\forall q\in L^1(X,\calX^0_s, Q_1)$, \
$\tilde\Lambda^s_t[q\eta]= q \tilde\Lambda^s_t[\eta] $, \ $0\leq s\leq t$, \ (i.e.
$\tilde\Lambda^s_t[q\eta](x)= q(x) \tilde\Lambda^s_t[\eta](x)$ a.s.).
\end{enumerate}
\end{assumption}

By points (3), (4) of Assumption \ref{Ass2} and (6) of Assumption \ref{Ass1}, one gets:
$\forall \sigma_s\in L^1\big(X,\calX_s^0, Q_1;\calT(\calH)\big)$, $0\leq s\leq t$,
\begin{equation}
\E_{Q_1}\big[ \tilde\Lambda^s_t[\sigma_s]\big|\calX^s_t\big]=
\tilde\Lambda^s_t\big[\E_{Q_1}[\sigma_s]\big].
\end{equation}
Here $\E_{Q_1}$ and $\E_{Q_1}[\bullet|\calX^s_t]$ are the classical expectation and
conditional expectation extended to operator-valued random variable.

Let us also define the \emph{evolution}
\begin{equation}
\calU(t,s)[\tau] := \E_{Q_1}\big[\tilde\Lambda^s_t[\tau]\big], \qquad \tau \in
\calT(\calH),\quad 0\leq s\leq t;
\end{equation}
$\calU(t,s)$ is a channel from $\calT(\calH)$ into $\calT(\calH)$. By points (2), (3), (4) of
Assumption \ref{Ass2}, for $0\leq r\leq s\leq t$, $\sigma_s\in L^1\big(X,\calX_s^0,
Q_1;\calT(\calH)\big)$, we get
\begin{equation}
\calU(t,s)\circ \calU(s,r)= \calU(t,r), \qquad
\E_{Q_1}\big[\tilde\Lambda^s_t[\sigma_s]\big|\calX^0_s\big] = \calU(t,s)[\sigma_s].
\end{equation}

The quantum continual measurements is represented by the operators $\tilde\Lambda ^s_t$, in
the sense that they give probabilities and state changes. If $\eta_0\in \calS(\calH)$ is the
initial state at time $0$ and $B\in \calX^0_t$ is any event involving the output in the
interval $(0,t)$, then $\int_B \Tr\{\tilde \Lambda^0_t[\eta_0](x)\}Q_1(\rmd x)$ is the
probability of the event $B$ and $\frac{\tilde\Lambda^0_t[\eta_0](x)}{ \Tr\{\tilde
\Lambda^0_t[\eta_0](x)\}}$ is the state at time $t$, conditional on the result $x$ (the
\emph{a posteriori} state). Instead, $\calU(t,0)[\eta_0]$ represents the state of the system
at time $t$, when the results of the measurement are not taken into account (the \emph{a
priori} state).

\section{The initial state and the measurement}
\subsection{Ensembles}
In quantum information theory, not only single states are used, but also families of quantum
states with a probability law on them, called ensembles. An \emph{ensemble} $\{\mu,\rho\}$ is
a probability measure $\mu(\rmd y)$ on some measurable space $(Y,\calY)$ together with a
random variable $\rho:Y\to \calS(\calH)$. Alternatively, an ensemble can be seen as a
quantum/classical state of the type described in Section \ref{qcalgebra}. Given an ensemble,
one can introduce an \emph{average state} $\overline \rho\in \calS(\calH)$
\begin{equation}
\overline \rho:= \E_{\mu} [\rho]=\int_Y \mu(\rmd y)\, \rho(y);
\end{equation}
the integrals involving trace class operators are always understood as Bochner integrals.
Finally, the average relative entropy of the states $\rho(y)$ with respect to $\overline \rho$
is called the ``$\chi$-quantity'' of the ensemble:
\begin{equation}\label{genchiq}
\chi\{\mu,\rho\}:= \int_Y \mu(\rmd y)\, S_\rmq \big(\rho(y)\ent \overline
\rho\big)=\E_\mu\left[S_\rmq \big(\rho\ent \overline \rho\big)\right].
\end{equation}
This new quantity plays an important role in the whole quantum information theory
\cite{Hol73,HolS04} and can be thought as a measure of some kind of quantum information stored
in the ensemble.

\subsection{The letter states}
Let us consider the typical setup of quantum communication theory.  A message is transmitted
by encoding the letters in some quantum states, which are possibly corrupted by a quantum
noisy channel; at the end of the channel the receiver attempts to decode the message by
performing measurements on the quantum system. So, one has an alphabet $A$ and the letters
$\alpha \in A$ are transmitted with some a priori probabilities $P_\rmi$. Each letter $\alpha$
is encoded in a quantum state and  we denote by $\rho_\rmi(\alpha)$ the state associated to
the letter $\alpha$ as it arrives to the receiver, after the passage through the transmission
channel. While it is usual to consider a finite alphabet, also general continuous parameter
spaces are acquiring importance \cite{YueO93,HolS04}.

\begin{assumption}\label{Ass3}
Let $(A,\calA,Q_0)$ be a probability space with $(A,\calA)$ standard Borel and let
$\sigma_\rmi$ be a normal state on $L^\infty\big( A,\calA,Q_0;\calL(\calH)\big)$.
\end{assumption}

Let us set
\begin{equation}
q_\rmi(\alpha):= \Tr\{\sigma_\rmi(\alpha)\}, \qquad
\rho_\rmi(\alpha):=\frac{\sigma_\rmi(\alpha)}{q_\rmi(\alpha)}\,, \qquad P_\rmi(\rmd \alpha):=
q_\rmi(\alpha)Q_0(\rmd\alpha);
\end{equation}
$q_\rmi$ is a probability density and $\{P_\rmi,\rho_\rmi\}$ is the initial \emph{ensemble}.
The average state and the $\chi$-quantity of the initial ensemble are
\begin{gather}
\eta_0 := \E_{Q_0} [\sigma_\rmi]=\int_A P_\rmi(\rmd \alpha)\, \rho_\rmi(\alpha),
\\
\chi\{P_\rmi,\rho_\rmi\}:= \int_A P_\rmi(\rmd \alpha)\, S_\rmq (\rho_\rmi(\alpha)\ent \eta_0).
\end{gather}
The quantity $\chi\{P_\rmi,\rho_\rmi\}$ is known also as Holevo capacity \cite{Hol73,HolS04}.

\subsection{Probabilities and states derived from $\eta_0$}

For $0\leq r \leq s \leq t$ we define:
\begin{equation}
\eta_t:= \calU(t,0)[\eta_0], \qquad \tilde \sigma^r_t:= \tilde \Lambda^r_t[\eta_r], \qquad
\tilde q^s_t:= \norm{\tilde \sigma^s_t}_1\,, \qquad \tilde \varrho^r_t:= \frac{\tilde
\sigma^r_t}{\tilde q^r_t}\,.
\end{equation}
Then, $\eta_t$ and $\varrho^r_t(x)$ are states on $\calL(\calH)$, $\tilde q^s_t$ is a state on
$L^\infty(X,\calX^0_t,Q_1)$ and $\tilde \sigma^r_t$ a state on
$L^\infty\big(X,\calX^0_t,Q_1;\calL(\calH)\big)$. We have also
\begin{equation}
\E_{Q_1}[\tilde q_t^r|\calX_s^0]= \tilde q_s^r\,, \qquad \E_{Q_1}[\tilde q_t^r|\calX_t^s]=
\tilde q_t^s\,.
\end{equation}
Moreover, there exists a unique probability $P_1$ on $(X,\calX)$ such that $P_1(\rmd x
)\big|_{\calX^0_t}= \tilde q^0_t(x) Q_1(\rmd x)$ for all $t\geq 0$.  Also $P_1(\rmd x
)\big|_{\calX^s_t}= \tilde q^s_t(x) Q_1(\rmd x)$ holds.

\subsection{The general setup}
It is useful to unify the initial distribution and the distribution of the measurement results
in a unique filtered probability space. Let us set:
\begin{subequations}
\begin{gather}
\Omega:= A\times X\,, \qquad \omega:=(\alpha,x), \qquad \pi_0(\omega):=\alpha\,, \qquad
\pi_1(\omega):=x\,,
\\
\sigma_0:= \sigma_\rmi\circ \pi_0\,, \quad q_0:= q_\rmi \circ \pi_0 = \norm{\sigma_0}_1\,,
\quad \rho_0 := \rho_\rmi\circ\pi_0 = \frac{\sigma_0}{\norm{\sigma_0}_1}\,,
\\
\calF := \calA\otimes \calX\,, \qquad  Q:= Q_0\otimes Q_1\,, \\ \calF_0:= \{B\times X : B\in
\calA\}, \qquad \calF^s_t:= \{A\times Y: Y\in \calX^s_t\} ,
\\
\calF_t:= \calF_0 \vee \calF^0_t =\sigma\{B\times Y : B\in \calA, Y\in\calX^0_t\}, .
\end{gather}
\end{subequations}

By defining \ $ \Lambda^s_t:= \openone\otimes\tilde\Lambda^s_t$, \ we extend \
$\tilde\Lambda^s_t$ \ to \ $L^1\big(\Omega,\calF_s,Q;\calT(\calH)\big)\simeq{}$  $
L^1(A,\calA,Q_0)\otimes L^1\big(X,\calX^0_s,Q_1;\calT(\calH)\big)$. \quad Similarly, we extend
\ \ $\calU(t,s)$ \ \ to \ \ $L^1\big(\Omega,\calF_s,Q; \calT(\calH)\big) \simeq
L^1(\Omega,\calF_s,Q)\otimes \calT(\calH)$. Let us also set:
\begin{subequations}
\begin{gather}
\sigma_t:= \Lambda^0_t[\sigma_0], \qquad \sigma^s_t:= \tilde \sigma^s_t\circ
\pi_1=\Lambda^s_t[\eta_s]\,, \qquad q_t:=  \norm{\sigma_t}_1\,, \\ q^s_t:=\tilde q^s_t \circ
\pi_1= \norm{\sigma^s_t}_1\,, \quad \rho_t := \frac{\sigma_t}{\norm{\sigma_t}_1}\,, \quad
\varrho^s_t:=\tilde \varrho^s_t\circ \pi_1= \frac{\sigma^s_t}{\norm{\sigma^s_t}_1}\,.
\end{gather}
\end{subequations}

In the computations of the following sections we shall need various properties of the
quantities we have just introduced; here we summarize such properties. Let $r,s,t$ be three
ordered times: $0\leq r \leq s \leq t$. Then, $\sigma_t$ and $\sigma^s_t$ are states on
$L^\infty\big(\Omega,\calF_t,Q;\calL(\calH)\big)$ and
\begin{subequations}
\begin{gather}
\E_Q[q_t|\calF_s]=q_s\,, \qquad\E_Q[q_t|\calF^s_t]=\E_Q[q^r_t|\calF^s_t]=q^s_t\,, \qquad
\\
\E_Q[q^r_t|\calF_s]=q^r_s\,,  \qquad
\E_Q[\sigma_t|\calF^s_t]=\E_Q[\sigma^r_t|\calF^s_t]=\sigma^s_t\,,
\\
\E_Q[\sigma_t|\calF_s]=\calU(t,s)[\sigma_s]\,, \qquad
\E_Q[\sigma^r_t|\calF_s]=\calU(t,s)[\sigma^r_s]\,,
\\
\E_Q[\sigma^s_t|\calF_s]=\eta_t\,, \qquad \eta_t= \E_Q[\sigma_t]\,, \qquad \sigma_t =
\Lambda^s_t[\sigma_s],
\\ \sigma^r_t = \Lambda^s_t[\sigma^r_s]\,,
\qquad \frac{\Lambda^s_t[\rho_s]} {\norm{\Lambda^s_t[\rho_s]}_1}=\rho_t\,, \qquad
\frac{\Lambda^s_t[\varrho^r_s]} {\norm{\Lambda^s_t[\varrho^r_s]}_1}=\varrho^r_t\,.
\end{gather}
\end{subequations}

We have that $\{q_t,t\geq 0\}$ is a non-negative, mean one, $Q$-martingale. Then, there exists
a unique probability $P$ on $(\Omega,\calF)$ such that $\forall t\geq 0$
\begin{equation}
P(\rmd \omega)\big|_{\calF_t}=q_t(\omega)Q(\rmd\omega).
\end{equation}
Moreover,
\begin{gather}
P(\rmd \alpha\times X)=P_\rmi(\rmd \alpha)\,, \qquad P(A\times \rmd x)=P_1(\rmd x)\,,
\\
P(\rmd\omega)\big|_{\calF^s_t}= q^s_t(\omega)Q(\rmd\omega)\,,\qquad \eta_t=\E_P[\rho_t]=
\calU(t,s)[\eta_s]\,.
\end{gather}

\section{Mutual entropies and informational bounds}

Here and in the following we shall have always $0\leq u\leq r \leq s \leq t$.

\subsection{The state $q_t$ and the classical information}

Let us consider the state $q_t$ and its marginals $\E_Q[q_t|\calF_r]=q_r$,
$\E_Q[q_t|\calF^r_t]=q^r_t$. Then, we can introduce the classical mutual entropy:
\begin{subequations}
\begin{equation}
S_\rmc(q_t \ent q_r q^r_t ) = \int_\Omega P(\rmd \omega) \, \log \frac {q_t(\omega)}
{q_r(\omega) q^r_t(\omega)}=:I_\rmc(r,t).
\end{equation}
Note that $I_\rmc(t,t)=0$. For $r=0$ we have the input/output classical information gain:
\begin{equation}
I_\rmc(0,t)= S_\rmc(q_t\ent q_\rmi \otimes \tilde q^0_t) \equiv \int_{A\times X} P(\rmd \alpha
\times \rmd x) \, \log \frac {q_t(\alpha,x)} {q_\rmi(\alpha) \tilde q^0_t(x)}\,.
\end{equation}
\end{subequations}

By applying the monotonicity theorem and the channel $\E_Q[\bullet|\calF_s]$ to the couple of
states $q_t$ and $q_r q^r_t$, we get
\begin{equation}
S_\rmc(q_t \ent q_r q^r_t )\geq S_\rmc\big(\E_Q[q_t|\calF_s]\ent \E_Q[q_r q^r_t|\calF_s]\big)
= S_\rmc(q_s \ent q_r q^r_s ),
\end{equation}
which becomes
\begin{equation}
I_\rmc(r,t)\geq I_\rmc (r,s)\,.
\end{equation}
The function $t\mapsto I_\rmc (s,t)$ is non decreasing.

\subsection{The state $\sigma_s$ and the main bound}

A useful quantity, with the meaning of a measure of the ``quantum information'' left in the a
posteriori states, is the mean $\chi$-quantity
\begin{equation}
\overline \chi(s,t):=\int_\Omega P(\rmd \omega)\, S_\rmq \big(\rho_t(\omega)\ent
\varrho_t^s(\omega)\big)= \E_P\left[S_\rmq \big(\rho_t\ent \varrho_t^s\big)\right].
\end{equation}
The interpretation as a mean $\chi$-quantity is due to the fact that $\overline
\chi(s,t)=\E_P\left[\E_P\left[S_\rmq \big(\rho_t\ent
\varrho_t^s\big)\big|\calF^s_t\right]\right]$. But by Eq.\ \eqref{genchiq} and
$\E_P[\rho_t|\calF^s_t]=\varrho^s_t$, $\E_P[\varrho^s_t|\calF^s_t]=\varrho^s_t$, we have that
$\E_P\left[S_\rmq \big(\rho_t\ent \varrho_t^s\big)\big|\calF^s_t\right]$ is a random
$\chi$-quantity. Note that
\begin{equation}
\overline \chi(t,t)= \int_\Omega P(\rmd \omega)\, S_\rmq (\rho_t(\omega)\ent \eta_t) =:
\chi\{P,\rho_t\}.
\end{equation}

Let us consider the state \ $\sigma_s$ \ and its marginals \
$\E_Q[\Tr\{\sigma_s\}|\calF_r]=q_r$, \ \ $\E_Q[\sigma_s|\calF^r_s]=\sigma^r_s$. Then, we have
the mutual entropy
\begin{equation}
S(\sigma_s\ent q_r\sigma^r_s)=I_\rmc(r,s)+ \overline \chi(r,s).
\end{equation}
For $r=s$ and for $r=s=0$ this equation reduces to
\begin{equation}
S(\sigma_s\ent q_s \eta_s)= \chi\{P,\rho_s\}, \qquad S(\sigma_0\ent q_0
\eta_0)=\chi\{P,\rho_0\}=\chi\{P_\rmi,\rho_\rmi\}.
\end{equation}

By applying the monotonicity theorem and the channel $\Lambda^s_t$ to the couple of states
$\sigma_s$ and $q_r\sigma^r_s$, we get
\begin{equation}
S(\sigma_s \ent q_r \sigma^r_s )\geq S\big( \Lambda^s_t[\sigma_s] \ent
\Lambda^s_t[q_r\sigma^r_s]\big) =S(\sigma_t \ent q_r \sigma^r_t ),
\end{equation}
which becomes
\begin{equation} \label{SWW2}
\overline \chi(r,s)-\overline \chi(r,t) \geq I_\rmc(r,t)- I_\rmc (r,s)\geq 0.
\end{equation}
Therefore, the function $t\mapsto \overline \chi(s,t)$ is non increasing.

For $r=s$ we get
\begin{equation}
S(\sigma_s\ent q_s \eta_s) \geq S(\sigma_t\ent q_s\sigma^s_t) ,
\end{equation}
which gives the upper bound for $I_\rmc$:
\begin{equation}\label{SWW1}
0\leq I_\rmc(s,t)\leq \chi\{P,\rho_s\} - \overline \chi(s,t).
\end{equation}
For $s=r=0$, it reduces to
\begin{equation}\label{SWW0}
0\leq  I_\rmc(0,t)\leq \chi\{P_\rmi,\rho_\rmi\} - \int_{A\times X} P(\rmd \alpha\times \rmd
x)\, S_\rmq \big(\rho_t(\alpha,x)\ent \tilde\varrho_t^0(x)\big).
\end{equation}
The bound \eqref{SWW0} is the translation in terms of continual measurements of the bound of
Section 3.3.4 of \cite{BarL05QP}, which in turn is a generalization of a bound by Schumacher,
Westmoreland and Wootters \cite{SchWW96}. Equation \eqref{SWW0} is a strengthening of the
Holevo bound \cite{Hol73} $I_\rmc(0,t)\leq \chi\{P_\rmi,\rho_\rmi\}$.

\subsection{Quantum information gain}
Let us consider now the \emph{quantum information gain} defined by the quantum entropy of the
pre-measurement state minus the mean entropy of the a posteriori states
\cite{Gro71,Lin72,Oza86}. It is a measure of the gain in purity (or loss, if negative) in
passing from the pre-measurement state to the post-measurement a posteriori states. In the
continual case, we can consider the quantum information gain in the time interval $(s,t)$ when
the system is prepared in the ensemble $\{P_\rmi,\rho_\rmi\}$ at time 0 or when it is prepared
in the state $\eta_r$ at time $r$:
\begin{subequations}
\begin{gather}
I_\rmq(s,t):= \int_\Omega P(\rmd \omega) \left[ S_\rmq\big(\rho_s(\omega)\big)-
S_\rmq\big(\rho_t(\omega)\big)\right],
\\
I_\rmq(r;s,t):= \int_\Omega P(\rmd \omega) \left[ S_\rmq\big(\varrho^r_s(\omega)\big)-
S_\rmq\big(\varrho^r_t(\omega)\big)\right].
\end{gather}
\end{subequations}
By this definition we have immediately
\begin{equation}\label{sum}
I_\rmq(r,t)= I_\rmq(r,s)+I_\rmq(s,t), \qquad I_\rmq(u;r,t)= I_\rmq(u;r,s)+I_\rmq(u;s,t).
\end{equation}

It has been proved \cite{Oza86} that the quantum information gain is positive for all initial
states if and only if the measurement sends pure initial states into pure a posteriori states.

As in the single time case \cite{BarL05os,BarL05QP,BarL05qic}, inequality \eqref{SWW2} can be
easily transformed into an inequality involving $I_\rmq$:
\begin{equation}\label{ineqIq}
I_\rmq(r;s,t) - I_\rmq(s,t) \geq I_\rmc(r,t)- I_\rmc (r,s)\geq 0.
\end{equation}

Let us take an initial ensemble made up of pure states:
$\rho_\rmi(\alpha)^2=\rho_\rmi(\alpha)$, $\forall \alpha\in A$. Let us assume that the
continual measurement preserve pure states: the states $\rho_t(\alpha,x)$ are pure for all
choices of $t, \alpha ,x$. Then, the von Neumann entropy of $\rho_t(\omega)$ vanishes and we
have $I_\rmq(s,t)=0$ for all choices of $s$ and $t$. From the second of Eqs.\ \eqref{sum} and
Eq.\ \eqref{ineqIq} we get
\begin{equation}
I_\rmq(u;r,t) - I_\rmq(u;r,s)= I_\rmq(u;s,t)\geq I_\rmc(u,t)- I_\rmc (u,s)\geq 0,
\end{equation}
i.e.\ the function $t\mapsto I_\rmq(u;r,t)$ is non decreasing for ``pure'' continual
measurements.

In particular, by taking $u=r=0$ we have
\begin{equation}
I_\rmq(0;0,t) = S_\rmq(\eta_0)- \int_X P_1(\rmd x) S_\rmq\big(\tilde \varrho^0_t(x)\big).
\end{equation}
For a continual measurement sending every pure initial state into pure a posteriori states,
$\forall \eta_0\in \calS(\calH)$ the quantum information gain $I_\rmq(0;0,t)$ is non negative,
non decreasing in time and with $I_\rmq(0;0,0)=0$.

\section*{Acknowledgments}
\noindent Work supported by the \emph{European Community's Human Potential Programme} under
contract HPRN-CT-2002-00279, QP-Applications, and by \emph{Istituto Na\-zio\-na\-le di Fisica
Nucleare}.

\end{document}